\documentclass[9pt,twocolumn,twoside]{opticajnl}
\usepackage{siunitx}
\usepackage{textcomp}
\usepackage{titlesec}
\usepackage[font=small,skip=0pt]{caption}
 \usepackage{layout}
\usepackage[toc,page]{appendix}
\usepackage{adjustbox}
\usepackage{subfigure}

\journal{opticajournal} 
\setboolean{shortarticle}{true}

\title{Optical Spring Tracking for Enhancing Quantum-Limited Interferometers}

\author[1]{Scott Aronson}
\author[1]{Ronald Pagano}
\author[2]{Torrey Cullen}
\author[3]{Garrett D. Cole}
\author[1]{Thomas Corbitt}

\affil[1]{Louisiana State University
Department of Physics \& Astronomy, Baton Rouge, LA 70803}
\affil[2]{LIGO Laboratory, California Institute of Technology, Pasadena, CA 91125}
\affil[3]{Thorlabs Crystalline Solutions, Santa Barbara, CA 93101}

\begin{document}
\begin{abstract}
Modern interferometers such as LIGO have achieved sensitivities limited by quantum noise, comprised of radiation pressure and shot noise. To mitigate this noise, a static system is employed that minimizes the quantum noise within the measurement band. However, since gravitational wave inspiral signals are a single frequency changing over time, only noise at the chirp frequency needs to be minimized. Here we demonstrate dynamically tracking a target signal using an optical spring, resulting in an increased signal to noise ratio (SNR). We report on a SNR increase by up to a factor of 40 when compared to a static configuration.


\end{abstract}

\maketitle

\section{Introduction}
Binary black hole and neutron star mergers are currently detected using laser interferometric observatories such as LIGO, Virgo and KAGRA \cite{LIGO,VIRGO,KAGRA,GW150914}. These gravitational wave signals are proportional to the time varying quadrupole moment of the merging system. As the separation distance between the two objects decreases due to the emission of gravitational waves (GWs), their rotational frequency and the resulting GW emission frequency increases in value until the merger. This increasing frequency is often referred to as the GW "chirp".

The LIGO and Virgo detectors have implemented higher power and frequency dependent squeezing to reduce their quantum noise \cite{PhysRevX.13.041021,Virgo_squeezing}. Frequency dependent squeezing is focused on the region of highest sensitivity, in the frequency range from 10-1000 Hz and originates from the injection of squeezed vacuum into a filter cavity. This cavity, which for LIGO is 300 m in length, produces the desired frequency dependence by rotating the squeezing quadrature of the light.  

In the current configuration, the quantum noise in LIGO is minimized around 100 Hz. This configuration is referred to as static, as the quantum noise is stationary with respect to time. Frequency dependent squeezing benefits from being able to reduce shot noise above 100 Hz without the cost of increasing radiation pressure noise \cite{PhysRevX.13.041021}.

With the future introduction of lower frequency band space-based GW detectors such as LISA \cite{Karsten}, we may acquire knowledge of incoming signals to the ground-based detector band with weeks to years of forewarning \cite{ligo_lisa,PhysRevD.99.124043}. Using this prior information of an incoming GW signal, we show that it is possible to dynamically tune the quantum noise to be minimized over the range of the chirp.

The optical spring effect occurs when a cavity is operated off resonance, or detuned, and has one or more mirrors susceptible to radiation pressure. The amount of circulating power and therefore radiation pressure depends on the total displacement from resonance. The optical field creates an effect analogous to a spring as the force from radiation pressure on the movable mirror is proportional to its position, similar to Hooke's law. By controlling the displacement from cavity resonance, the optical spring's resonant frequency can be actively controlled.

The use of an optical spring has been shown to provide different types of noise suppression \cite{backaction_cancellation,cullensubsql,passive_pow_stab}. In this letter we report a proof of principle experiment to dynamically control an optical spring \cite{13} in order to track a target signal over time \cite{Dyn_tune_inter,Dyn_tine_time,Dyn_tune_experiment}. This dynamic tracking increases the SNR of the measurement when compared to a static configuration, improving the sensitivity by up to a factor of 40 in our experiment. For interferometer configurations such as LIGO, the optical spring effect can be achieved through the detuning of the signal recycling cavity \cite{PhysRevD.99.124043,Chen}.


To understand the noise with the optical spring present we start with an expression of the circulating power within the cavity \cite{cripe_thesis}. This derivation assumes an instantaneous response time of the cavity to changes in the circulating power from changes in detuning.

\begin{equation}
P_{c}=\frac{P_{0}}{1+\delta^{2}}
\label{eq:P_circ}
\end{equation}

The maximum on resonance circulating power is given by $P_{0}$, and the detuning $\delta$ is in units of linewidths of the cavity. The optical spring constant $k_{OS}$ in the static limit is given by \cite{Corbitt_thesis},

\begin{equation}
k_{O S} =\frac{2}{c} \frac{d P_{c}}{d x} 
 =\frac{2}{c} \frac{d P_{c}}{d \delta} \frac{d \delta}{d x}. 
\label{eq:optical_spring_constant}
\end{equation}

\noindent The displacement from resonance ,$x$, can be written in terms of the detuning as \cite{cripe_thesis}:
\begin{equation}
    x = \frac{\delta \lambda T}{8\pi},
    \label{eq:x}
\end{equation}

\noindent where T is the total losses within the cavity and $\lambda$ is the laser wavelength.
This yields a final expression for $k_{OS}$,

\begin{equation}
    k_{OS} =\frac{-32 \pi \delta P_{c}}{\lambda c T\left(1+\delta^{2}\right)}. 
    \label{K_OS}
\end{equation}


At the optical spring's resonant frequency mirror motion is amplified by the optical spring's quality factor, $Q_{OS}$. $Q_{OS}$ results from the non-instantaneous response time of the cavity acting as a damping force resulting in a finite amplification of the mirror motion. It is beneficial to remove the effect of this amplification to recover the expected mirror motion without an optical spring present. In this "free mass regime", the signal amplification from $Q_{OS}$ is divided out and any noise external to the cavity such as shot noise at the optical spring frequency is effectively suppressed.




The radiation pressure in the free mass regime due to the shot noise fluctuations in the circulating beam is calculated by,

\begin{equation}
\begin{aligned}
x_{r p} & =\frac{1}{m \Omega^{2}} \frac{2 P_{c}}{c} \sqrt{\frac{2 h f}{P_{i n}}} \\
& =\frac{1}{m \Omega^{2}} \frac{2 P_{c}}{c} \sqrt{\frac{8 h f}{P_{c} T\left(1+\delta^{2}\right)}}.
\end{aligned}
\end{equation}

\noindent where $\Omega$ is the angular frequency, $h$ is Plank's constant, $P_{in}$ is the power incident on the cavity, c is the speed of light, and $f$ is the laser frequency.

A standard reference for the performance of interferometric measurement devices is the standard quantum limit (SQL) \cite{Braginsky547}. This defines the minimum noise in the uncorrelated sum of the quantum radiation pressure and shot noise. The SQL is not a hard limit however. It has been shown one method to achieve a measurement below the free mass SQL is through the use of an optical spring \cite{cullensubsql}.


To evaluate the performance at the optical spring frequency we follow the derivation from \cite{cullensubsql}. First we calculate the free mass SQL via the following equation,

\begin{equation}
    x_{S Q L} =\sqrt{\frac{2 \hbar}{m \Omega^{2}}}.
\end{equation}

We then set $\Omega$ = $\Omega_{OS}$ and divide the radiation pressure $x_{rp}$ by $x_{S Q L}$. Here $\Omega_{OS} = \sqrt{\frac{k_{OS}+k_m}{m}}$, where $k_m$ is the mechanical spring constant and $m$ is the mass of the movable mirror. Since $k_m \ll k_{OS}$, we can approximate $\Omega_{OS} \approx \sqrt{\frac{k_{OS}}{m}}$.

By recalling $k_{OS}$ from Eq. \ref{K_OS} we find,
\begin{equation}
\begin{aligned}
\left. \frac{x_{r p}}{x_{S Q L}}\right|_{\Omega_{OS}} &=\frac{1}{c} \sqrt{\frac{32 \pi h f P_{c}}{m \left(\sqrt{\frac{k_{OS}}{m}} \right)^2 T\left(1+\delta^{2}\right)}}\\
&=\sqrt{\frac{1}{-\delta}} .
\end{aligned}
\label{eq:1/delta}
\end{equation}

The resulting noise level is predicted to be below the standard quantum limit for detunings less than -1 \cite{cullensubsql}. The above result is only valid for detunings $\delta<0$ where a positive optical spring constant is present and $Q_{OS}$ is sufficiently large. For $\delta>0$ no optical spring resonance occurs and there is no suppression of the noise external to the cavity. Equation \ref{eq:1/delta} predicts that as we sweep the optical spring to follow the target signal frequency, we surpass the SQL by a factor of $\sqrt{-\delta}$.

\section{Experimental Setup}\label{Section 2}

\begin{figure}[h!]
\centering
\includegraphics[width=\linewidth]{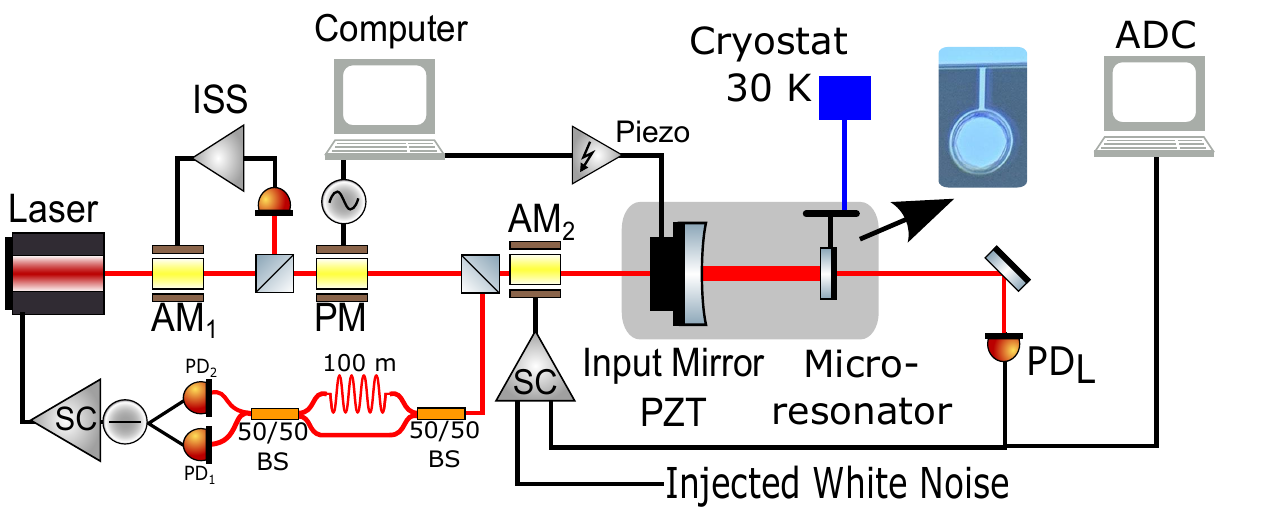}
\caption{Simplified Experimental Setup of optical spring tracking. The computer is used to inject a signal to the phase modulator (PM). The same computer also controls the detuning of the cavity via a piezo mount to track the signal frequency via a dynamically tunable optical spring. White noise is injected into the servo controller (SC) stabilizing the cavity.}
\label{fig:setup}
\end{figure}

The experimental setup for the optical spring tracking can be found in Fig. \ref{fig:setup}. The experiment consists of a 1064 nm Nd:YAG nonplanar ring oscillator (NPRO) laser. The laser intensity is actively stabilized via an intensity stabilization servo (ISS) and its frequency is controlled through a 100 m delay line interferometer. 

The optical cavity is stored in a vacuum chamber at $10^{-8}$ torr and the micro-resonator is kept at cryogenic temperatures of around 30 K. The micro-resonator acting as the output mirror is an AlGaAs Bragg reflector with an effective mass of $\sim$50 ng, supported by a thin GaAs cantilever \cite{cripe_thesis}. The resonator has a fundamental mechanical resonance frequency of 876 Hz and quality factor, $Q_{m}$, of 25000 $\pm$ 2200. The input stationary mirror has a radius of curvature of 1 cm and forms a stable cavity 1 cm in length. Within the vacuum chamber is a seismic isolation platform which strongly suppresses by 50 dB all motion above 100 Hz \cite{cripe_thesis}.

For an optomechanical cavity to be dynamically stable, there must be a positive optical spring constant along with a positive damping coefficient \cite{Singh_PRL,17}. In order to introduce a positive damping coefficient, an amplitude modulator (AM$_{2}$ in Fig. \ref{fig:setup}) is employed to stabilize the system.

A sinusoidal signal of varying frequency is injected into the laser beam via a phase modulator (PM). The signal to the PM was chosen to be constant in magnitude and linearly increase in frequency from 40 kHz to 100 kHz over 10 seconds. 
The same computer which controls this signal injected to the phase modulator is also used to simultaneously vary the detuning of the cavity via a piezo mount on the input mirror. The piezo offset is calibrated in order to generate a desired detuning and thus optical spring frequency. To track the signal, the optical spring frequency is controlled to follow the frequency sent to the PM over time. 


White noise is injected after PD$_{L}$ to simulate additional shot noise and enhance the effect of the optical spring noise suppression. Four measurements were performed to show the performance of the optical spring tracking system. First the detuning is swept with no signal present to determine the background. Then the detuning and the signal are swept together to establish a signal present over the background. To compare the performance to a static configuration, we fix the optical spring at 70 kHz and undertake the same procedure.

The data is saved via an ADC at 500 kHz sampling rate over 10 seconds. The time-series are then processed into a spectrogram and averaged in time using a moving average with a window of 100 samples to remove excess transient noise in the measurement.

\section{Numerical Simulations} \label{section2}
In order to understand the total noise contributions of the system we employ a numerical model utilizing the two photon formalism \cite{Corbitt_mathematical}. 
This numerical simulation accounts for losses and the non-instantaneous response time of the cavity. For the purpose of this letter, we neglect the contributions of classical noise sources such as thermal noise and focus solely on the intrinsic quantum noise of the system. Such quantum limited interferometric systems have been demonstrated previously \cite{Cripe_QRPN,cullensubsql}.  


\begin{figure}[h!]
\centering
\includegraphics[width=0.9\linewidth]{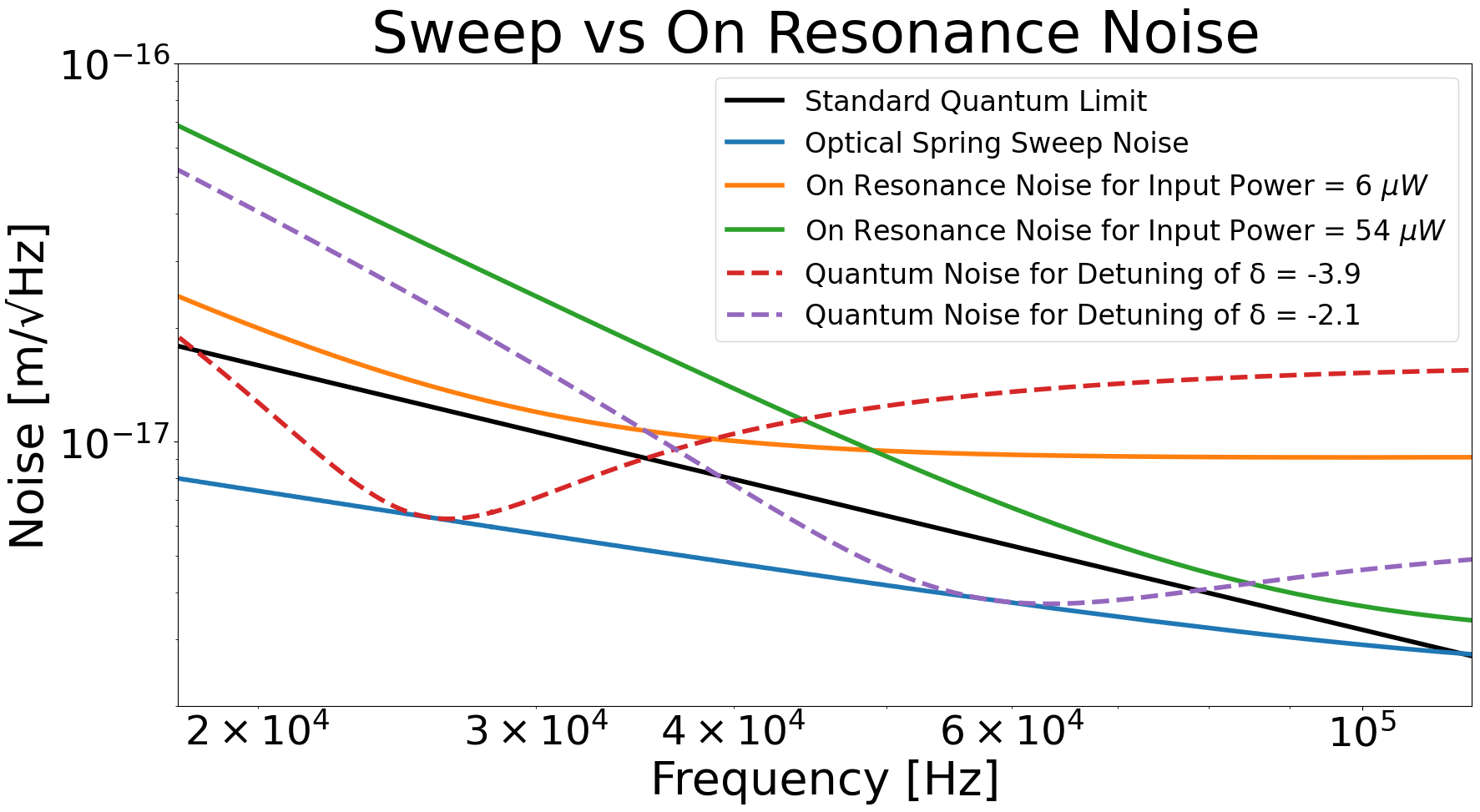}
\caption{Effective Sweep Noise compared to the Standard Quantum Limit (SQL) and the On Resonance Noise for two input powers. The Sweep Noise is predicted to be below the on resonance noise and is sub-SQL in the free-mass regime. The Optical Spring Sweep Noise curve is constructed from individual detuning curves (shown as dotted lines).}
\label{fig:off vs on res}
\end{figure}

The key experimental parameters we feed into the model are the length of the cavity L = 0.01 m, the input power $P_{in} = 0.5$ mW, cavity losses of 220 ppm, a stationary input mirror of transmission T = 50 ppm, and a movable cantilever with T = 450 ppm and a mass of m = 50 ng. With these parameters, the cavity is dominated by radiation pressure below 100 kHz. 

The code is then iterated over 100 detunings starting at a weak optical spring with $\delta = -5$ and ending with the maximum optical spring at $\delta = -\frac{1}{\sqrt{3}}$. For each detuning the minimum in the quantum noise is recorded as 'Optical Spring Sweep Noise'. The resulting curve is plotted in Fig. \ref{fig:off vs on res}. For all modeled frequencies below 100 kHz the Sweep Noise has lower noise than the free mass SQL.

As the quantum radiation pressure in our experimental setup is the dominant source of noise \cite{Cripe_QRPN}, operating the system on resonance in practice is nontrivial. For the experimental results discussed in Section \ref{Section 4} we compare the noise at a single stationary optical spring frequency to the effective noise from sweeping the optical spring/detuning. The Sweep Noise is predicted to outperform the static configuration at all but a narrow band where the two optical spring frequencies overlap. 

\begin{figure}[h!]
\centering
\includegraphics[width=0.9\linewidth]{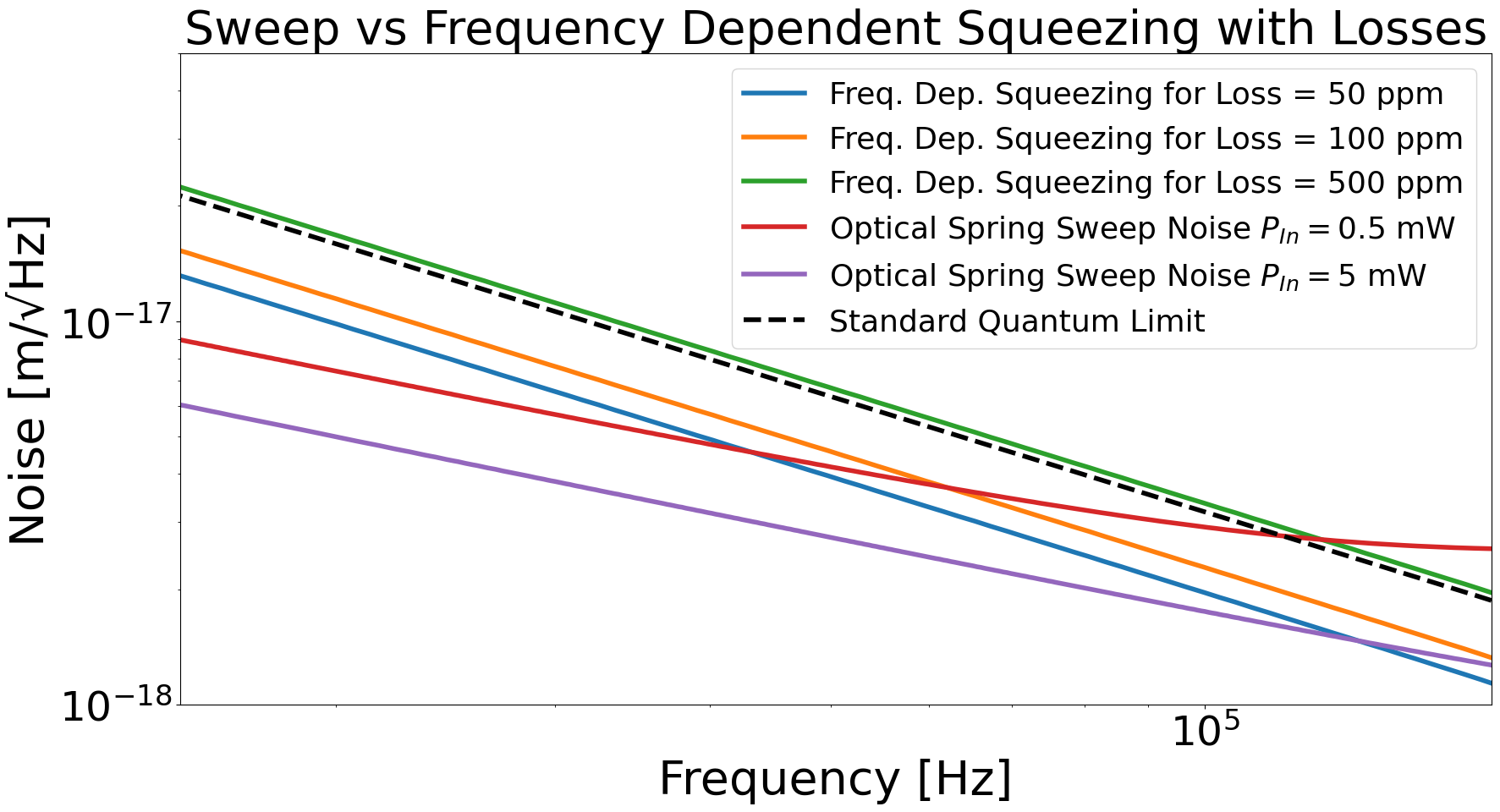}
\caption{Optical Spring Sweep Noise curves compared to modeled on resonance frequency dependent squeezing with 4.6 dB of output squeezing. The frequency dependent curves are a collection of the minimum noise as a function of input power.
As optical losses are introduced the frequency dependent measurements lose sensitivity. The optical spring sweep noise can be reduced through an increase in input power.}
\label{fig:freq_dep_loss}
\end{figure}

The frequency dependent squeezing currently implemented in LIGO and Virgo is created by the injection of a squeezed vacuum state into the detector. Any optical losses contaminate the state through the introduction of unsqueezed vacuum \cite{PhysRevLett.124.171102}. Detectors must maintain very low optical losses to see the full benefit of squeezing. Optical spring tracking does not suffer from the same limitation of optical losses disrupting the measurement because the motion of the mirror is amplified. With frequency dependent squeezing, the motion remains constant, while the noise is reduced, which makes it more sensitive to losses. The effect of increased losses with frequency dependent squeezing compared to the optical spring Sweep Noise can be found in Fig. \ref{fig:freq_dep_loss}. Each frequency dependent squeezing curve is constructed from the minimum noise for each frequency when sweeping the input power.

\begin{figure*}[h!]
\centering
\footnotesize
\includegraphics[width=4.75in]{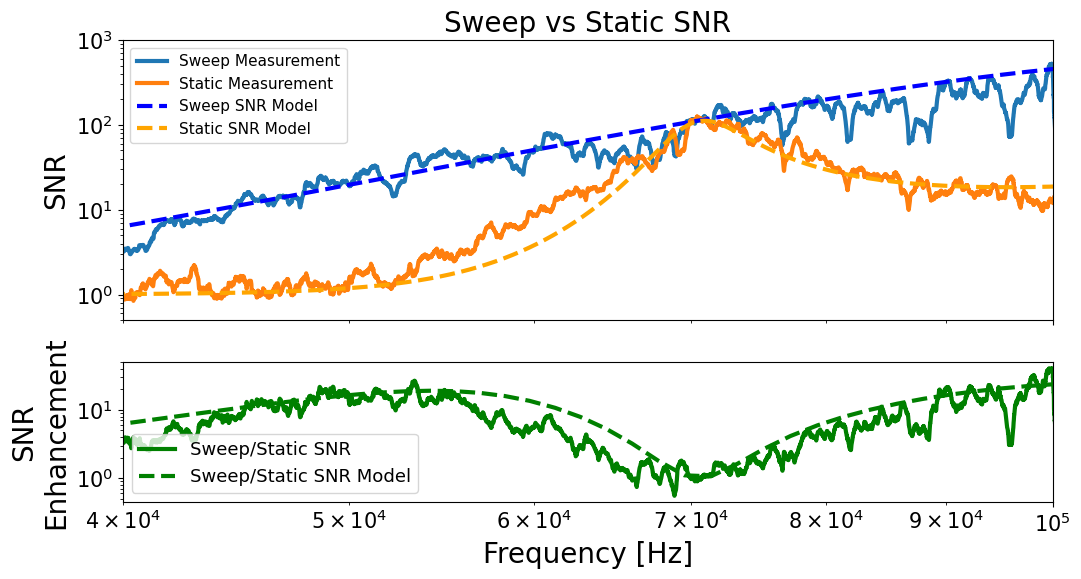}
\captionsetup{width=\linewidth}
\caption{Comparison of the injected signal SNR for sweep vs static configurations. The sweep configuration outperforms the static configuration for all but 70 kHz, where the optical spring frequencies were in proximity. The lower plot shows the ratio of the sweep SNR to the static configuration. The SNR enhancement has an average of 8.5 across the band, reaching a maximum value of 40.5 at 100 kHz.}
\label{fig:SNR_compare}
\end{figure*}


Currently in the LIGO interferometers, 17 dB of squeezing is generated while the extrinsic optical losses result in just 4-5.8 dB of usable squeezing \cite{PhysRevX.13.041021}. The minimum cavity loss of 50 ppm in Fig. \ref{fig:freq_dep_loss} was chosen by modeling 17 dB of injected squeezing into our system and setting the loss such that the output squeezing of 4.6 dB was within the range of LIGO's measured squeezing.

For cases where high optical losses cannot be avoided, optical spring tracking provides an alternative technique for quantum noise reduction. This method, unlike frequency dependent squeezing, does not require the use of a filter cavity.

\section{Results}\label{Section 4}

Using the background spectrogram and signal spectrogram, the SNR is calculated by evaluating the difference between the signal and the background, then dividing this difference by the background. With the SNR of both the sweep and static configurations calculated they are then directly compared. For each time slice of the SNR spectrogram the maximum value is recorded along the signal’s frequency. This gives a measure of the signal extraction for each configuration as a function of frequency, as seen in Fig. \ref{fig:SNR_compare}. There is a clear increase in the SNR of the sweep configuration vs the static configuration for all frequencies except 70 kHz where their optical spring frequencies overlap.


Limitations of this technique include the need for prior knowledge of the time and morphology of an incoming signal. For our current implementation, there is also a limitation stemming from the bandwidth of the piezo response. In practice something like an AOM or EOM could be implemented to directly modulate the laser frequency, granting more dynamic range.


The frequency range over which the optical spring can be easily positioned depends strongly on the configuration such as input power and detuning. So far we have only considered the optomechanical resonance, but there is also an optical resonance. The optical resonance can also be controlled via the detuning and is higher in frequency than the optomechanical resonance. While the optomechanical resonance is more suited to measure earlier in a GW inspiral, the optical resonance may be used to broaden the measurement to frequencies up to and including binary black hole coalescence \cite{PhysRevD.99.124043,ligo_minsweep}. 

The optical spring control could be further enhanced by the introduction of optical parametric amplification \cite{Dyn_tune_engineering,Dyn_tune_experiment}. This amplifier can modify the frequency of the optical spring providing an alternative method of tuning the resonators response.

\section{Summary}

We have predicted that by dynamically tracking the optical spring frequency to a known signal, a measurement can be performed with lower noise than on resonance cases and surpasses the free mass standard quantum limit (SQL). When experimentally comparing the dynamic optical spring tracking to a static optical spring configuration, there is an average increase in SNR by a factor of 8.5 with a maximal increase by a factor of 40 at 100 kHz.

Optical spring tracking can be utilized in any quantum limited interferometric measurement system to measure a known signal changing in frequency over time. This method could serve as a complement \cite{ligo_minsweep} or an alternative to frequency dependent squeezing for systems, particularly those limited by optical losses. If desired, the method can track a single event, filtering out concurrent signals. 

This technique could be used in the future for ground based gravitational wave interferometers which have foreknowledge of a signal’s arrival time and morphology. Such information may be ascertainable from planned projects such as LISA whose measurement band sits below LIGO. 

\subsection*{Funding}
\small{\noindent S. Aronson, R. Pagano, and T. Corbitt are supported by the NSF grant PHY-2110455.}
\subsection*{Disclosures}
\small{The authors declare no conflicts of interest.}

\subsection*{Data availability}
\small{Data underlying the results presented in this paper are not publicly available at this time but may be obtained from the authors upon reasonable request.}

\bibliography{main}

\bibliographyfullrefs{main}

\end{document}